**Size-selective nanoparticle growth on few-layer graphene films**

Zhengtang Luo,[†] Luke A. Somers,[†] Yaping Dan,[‡] Thomas Ly,[†] Nicholas J. Kybert,[†] E. J. Mele,[†] A.T. Charlie Johnson[†] [*]

[†]*Department of Physics and Astronomy, University of Pennsylvania, Philadelphia, PA 19104*

[‡]*Department of Electrical and Systems Engineering, University of Pennsylvania, Philadelphia, PA 19104*

**ABSTRACT**

We observe that gold atoms deposited by physical vapor deposition onto few layer graphenes condense upon annealing to form nanoparticles with an average diameter that is determined by the graphene film thickness. The data are well described by a theoretical model in which the electrostatic interactions arising from charge transfer between the graphene and the gold particle limit the size of the growing nanoparticles. The model predicts a nanoparticle size distribution characterized by a mean diameter $\bar{D}$ that follows a $\bar{D} \propto m^{1/3}$ scaling law where *m* is the number of carbon layers in the few layer graphene film.

Graphene, a honeycomb sheet of $sp^2$-bonded carbon atoms, has attracted immense attention since the first demonstration of reliable methods to rapidly isolate and identify it.[1] More recently it has become clear that graphene multilayers or "few layer graphene" (FLG) films actually represent a *family* of materials with properties that vary systematically with the number of carbon layers. As an example, prior theoretical work considered the electronic structure of FLG and the effect of a substrate on the spectrum.[2-7] More recently, electrostatic force microscopy was used to measure very significant variations of the surface potential with thickness for FLG films on an oxidized silicon substrate,[8] which were found to be in excellent agreement with a nonlinear Thomas-Fermi theory for the FLG charge carriers. Here we report on another important effect of FLG film thickness, namely its impact on the diameter of nanoscale metal particles formed on the FLG surface.

Thick graphite samples are widely used as substrates for the preparation of free-standing metal nanoparticles due to their chemical inertness, low binding energies for adsorbed atoms, low energy barrier for lateral diffusion, and suitability for various electron microscopies, in particular for scanning tunneling microscopy.[9, 10] Here we study the formation of nanoparticles on FLG films, and observe that the size of the growing nanoparticles is controlled by film thickness. We find that mean particle diameter $\bar{D}$ increases with the layer count of the graphene $m$, as does the variance of the distribution of particle radii. The observations are compared with the predictions of a theoretical model where long-range electrostatic interactions resulting from charge exchange between graphene and the nanoparticle limit the growth of large nanoparticles. The nanoparticle diameter is determined by a competition between its electrostatic energy and the surface tension of the particle. The model predicts that $\bar{D}$ increases as $\bar{D} \propto m^{1/3}$, while the variance of the radius distribution grows as $\sigma_R^2 \sim m$. The power law dependence of $\bar{D}$ occurs as a consequence of the unconventional charge screening in the graphene film.[8]

We find very good agreement between the data and the predictions of the model. Specifically, we fit the average diameter data to a power law $\bar{D} = Am^{\nu}$, as suggested by the model. The best-fit values for the exponent and prefactor are $\nu = 0.33 \pm 0.06$ and $A = 6.46 \pm 0.68$ nm, respectively, which compare well with the theoretical predictions $\nu = 1/3$ and $A = 5.9$ nm. Good agreement is also found for the variance data (see below).

Graphene samples were obtained by micromechanical cleavage of Kish graphite and placed on 300nm oxidized substrates.[1] The number of graphene layers was determined by cross-corroboration between interference induced color shift in optical microscope and atomic force microscope (AFM), and further confirmed by their Raman spectra[11, 12] (see Fig. S1 of the Supplemental Information). Sample cleanliness was found to be an important factor in obtaining reproducible results, so the as-prepared graphene samples were annealed at 400°C in forming gas (10% hydrogen, 90% ultrahigh purity argon gas) for 5 hours to remove adsorbates and contamination. Thermal evaporation of gold was performed using an alumina-coated tungsten boat at vacuum pressure of $5\times10^{-7}$ Torr, while the deposition flux (0.05 nm/s) and final thickness (0.3 nm) were monitored with a quartz crystal microbalance. The sample was then loaded in a quartz tube chamber and annealed at 400°C for 3 hours in a flow of forming gas. This allowed nanoparticles to form and reach an equilibrium configuration, as evidenced by the fact that no further particle morphology change was observed when such a sample was annealed for 4 additional hours.

Figures 1(a) and 1(b) are high-resolution scanning electron micrographs of identically prepared samples of Au nanoparticles that differ only in the thickness of the FLG film (~ 700 layers for Fig 1a and a single layer for Fig 1b; see Supplemental Information). Striking differences are observed in the size and shape of the gold nanoparticles between thick and thin films.

For thick FLG films on oxidized silicon substrates, we find a nanoparticle morphology (Fig 1a) identical to that reported for gold nanoparticles formed on bulk HOPG after vacuum annealing,[13] indicating that the presence of the silicon substrate is negligible in this configuration. There are a comparatively small number of large nanoparticles (diameter ~ 100-300 nm) whose facets are oriented along directions rotated by multiples of 60 degrees, as indicated in Fig 1a and analyzed in the inset. This is similar to what is observed in the epitaxial growth of micron-sized Au particles on graphite [14] where the {111} face of the metal grows on the {0001} face of graphite. In this case, particle growth is governed by the well-studied Ostwald ripening process, where large particles grow at the expense of the smaller particles, because of the latter's proportionately higher surface energy.[15]

For Au deposited onto single- and few-layer graphene films (Fig. 1b), however, the nanoparticle morphology after annealing is radically different. The FLG film surface is now decorated by a large number of comparatively small nanoparticles (diameter 5-20 nm) with no evidence of faceting. This configuration represents the equilibrium state of the system since no significant further size changes are observed after several additional hours of annealing.

We find that the average nanoparticle diameter, measured by HRSEM, and the statistical variation in this quantity are strongly dependent on the thickness of the FLG film, with the smallest particles and narrowest distribution observed on single layer graphene (see, for example, Fig. 2a which shows particles grown on FLG films of 1, 2, and 3 layers thickness). These quantities are plotted in Fig. 2b and its inset, along with power law fits that are motivated by an analysis of the particle free energy as discussed below. Additional detail and AFM and SEM images are provided in Figs. S5-12 in the Supplemental Information. We also find that the average diameter-to-height ratio for nanoparticles grown on FLG films is approximately 5, so that the particles grow as relatively "flat" islands. This is inferred from Fig. 3, which shows the height (measured by AFM) versus diameter for nanoparticles grown on a FLG film. For this plot, the particle height is measured by AFM (accuracy of 0.1 nm), while the lateral dimension of the same particle is measured by HRSEM in order to avoid tip convolution broadening of 5-10 nm associated with the AFM tips used (see Fig. S4 in Supplemental Information for details). Thus, nanoparticles grown on FLG films of fewer than 20 layers grow as relatively flat cylinders, with an average diameter controlled by the FLG thickness.

Our observation that the equilibrium configuration of metal nanoparticles formed on thin FLG films differs strongly from that of particles grown on thick graphene films suggests that the interactions that control particle growth differ sharply as well. The large particles on thick graphene apparently grow by the Ostwald ripening process that is determined by the interplay of bulk cohesive energy and surface interactions. The different morphology of nanoparticles on FLG indicates the impact of a new long-range interaction whose strength grows with decreasing FLG thickness. Prior work[8, 16] suggests the likelihood of an electrostatic interaction whose strength is thickness dependent. Indeed, we find that the data are well described by a model where electrostatics play a critical role in the size selection, and so we examine this interaction in detail before constructing a model for the size-dependent total energy of the particle.

Consider the effect of the work function mismatch $\Delta\Phi$ between graphene (~4.66 eV[17]) and gold (5.1-5.47 eV). When the gold nanoparticle is adsorbed on graphene there is a transfer of electrons from the graphene to the nanoparticles, leaving the particle negatively charged with an interfacial charge density $\sigma_0$. Each donated charge produces a dipole of strength $p = es$ where $s$ parameterizes the effective width of the dipole layer. This width is determined by the physical penetration depth of the interfacial charge density on the gold particle and the depth of the screening charge in the graphene film. Since few layer graphene films screen poorly, $s$ is effectively limited by the penetration depth in the graphene. An areal dipole density $\tau = \Delta \Phi/4 \pi e^2 s$ is then required to equilibrate the Fermi energies of the metal nanoparticle and the graphene.

The interfacial dipoles are all equal and oriented normal to the surface with the positive end outward, leading to a (positive) electrostatic energy for the droplet

$$U_d = \frac{\Delta\Phi^2}{32\pi^2 e^2} \int \frac{d^2 r d^2 r'}{|r - r'|^3}, \quad (1)$$

where the integral is taken over the cross sectional area of the droplet. This integral is formally divergent because of the small distance behavior of the integrand. The physical energy is finite since the integral is cutoff at small length scales by the spatial extent of the dipole. Below this scale (1) reverts to a nonsingular near field form that makes a contribution to the bulk energy, proportional to the droplet volume. Thus, by integrating (1) for $r > d$, where $d$ is the thickness of the graphene film, we obtain the long-range contribution to the energy that shows a new scaling with droplet size.

The interaction energy can be equivalently expressed, using a Fourier transform, as

$$U_d = \frac{\Delta\Phi^2}{32\pi^2 e^2} \int d^2 q V(q) |S(q)|^2 \quad (2)$$

where $S(q) = 2\pi \int_0^R r J_0(qr)\, dr$ is the form factor for a circular island with radius $R$. By cutting off the interaction range at the graphene film thickness we find that $V(q \to 0) = 1/d$ and thus the long range electrostatic interaction energy becomes

$$U_d = \frac{\pi^2 \Delta\Phi^2 R^4 \tau}{8e^2 d} \qquad (3)$$

The areal density $\tau$ appearing in this expression depends on the Fermi energy mismatch $\Delta\Phi$, and it can be deduced from a Thomas Fermi model for interlayer screening developed by us previously in Ref. 8 which gives $\tau = C\Delta\Phi^{3/2}$ with $C = \sqrt{3/8}\left(1/\hbar v_F \sqrt{e^2 c}\right)$, where $c = 0.34$ nm is the interlayer spacing in graphite. Thus, eliminating the areal dipole density, the electrostatic energy reads

$$U_d = \frac{\sqrt{3}}{8^{3/2}} \frac{\pi^2 R^4}{\hbar v_F d e^3 \sqrt{c}} \Delta\Phi^{7/2} = \frac{\Gamma}{m} R^4 \qquad (4)$$

where $m = d/c$ is the layer count of the graphene. This is a useful result, since it demonstrates that the long range electrostatic energy scales as the square of the droplet area, inversely with the film thickness and vanishes if there is no charge transfer because of matched work functions. Assuming a work function difference $\Delta\Phi = 0.5$ eV, we find a value for $\Gamma = 0.61$ eV/nm$^4$.

This electrostatic interaction can now be included to complete a model for the total energy of a condensed droplet. This energy has contributions from the particle surface area $A$ and its volume $V$ as well as the electrostatic contribution

$$U = \gamma A + uV + \frac{\Gamma}{m} R^4 \qquad (5)$$

For a cylindrical particle with radius $R$ and height $h$ we have $A = 2\pi R^2 + 2\pi Rh$, $V = \pi R^2 h$, $\gamma$ represents the surface tension (energy/area; from the literature[18], we find $\gamma = 9.6$ eV/nm$^2$) and $u$ (<0) is the bulk cohesive energy per unit volume. In principle, one needs to treat separately the energies of the free and contacted (bottom) surfaces of the droplet, but this detail is unnecessary in the following scaling analysis. We note that in the absence of the electrostatic term, the model exhibits the familiar critical droplet phenomenon at a critical radius $R_C = \gamma h / (|u| h - 2\gamma)$ below which the particles collapse and above which the particles grow without bound. The long-range electrostatic term then has the effect of limiting the growth of large droplets. It is useful to express the result by computing the chemical potential

$$n\mu = \left(\frac{1}{2\pi R h}\right)\frac{\partial U}{\partial R} = u' + \frac{\gamma}{R} + \frac{2\Gamma R^2}{\pi m h} \qquad (6)$$

where $n$ is the atomic number density (for gold, $n = 59/\mathrm{nm}^3$ [19]), and $u' = u + \Gamma / h$ gives the (constant) volume dependent term in the droplet energy.

Figure 4(a) illustrates a plot of the phenomenological free energy as a function of disk diameter for various thickness of graphene, as obtained from Eq. 6. The total droplet energy increases linearly for small-diameter particles because its free energy is dominated by the line tension on the boundary. For intermediate particle diameters, the negative volume energy makes the overall energy negative. Electrostatic energy dominates the large R behavior of the energy and prevents a runaway growth of the droplet. Figure 4(b) illustrates chemical potential of nanoparticles grown on different number of graphene layers, calculated using Eq 6. The chemical potential diverges $\propto$ 1/R for small R from the surface tension term and it increases $\propto$ $R^2$ at large R because of the dipolar interactions. The chemical potential curve sharply rises for small diameters, with a gradual increase for large diameters. The curve has a broader minimum for thicker graphene, which predicts a broader diameter distribution for such flakes.

In equilibrium the system forms droplets at the average radius that minimizes the chemical potential $\mu(R)$. Since the potential of Fig. 4(a) is concave downward, in practice this implies a microphase separation of the adsorbed gold into a low density (uncondensed) phase and droplet (condensed) phase at an average radius corresponding to this minimum. From Eq. (6)

we can predict that for an ensemble of droplets in thermal equilibrium, the average diameter follows:

$$\bar{D} = 2\bar{R} = \left(\frac{2\pi\gamma mh}{\Gamma}\right)^{1/3} = 5.9m^{1/3} \text{ nm} \qquad (7)$$

The curvature of $\mu(R)$ around its minimum value is $\mu'' = 12\Gamma / \pi hmn$. Thus for an equilibrium ensemble of droplets that are formed at the annealing temperature $T$, one predicts a distribution of droplet radii with a variance that satisfies:

$$\begin{aligned} &\frac{1}{2}\mu''\left(\langle R^2\rangle - \langle R\rangle^2\right) = \frac{kT}{2} \\ &\sigma_R^2 = R^2 - \bar{R}^2 = \frac{\pi hmnkT}{12\Gamma} = 1.6m \text{ nm}^2 \end{aligned} \qquad (8)$$

where $\langle\ \rangle$ denotes a thermal average. To determine the numerical prefactors in Eq. 7-8, we use the values of $\gamma$, $\Gamma$, and $n$ given previously, along with a value of $h = 2$ nm taken from the data in Fig. 3.

These theoretical predictions motivate the curve fits to the data presented in Fig. 2(b) and its inset. The particle sizes and standard deviations increase with graphene thickness. Fitting the average diameter data to a power-law yields an exponent of 0.331 ± 0.061 and a prefactor of 6.46 ± 0.68 nm, in good agreement with the theoretical predictions of 1/3 and 5.9 nm, respectively. The measured variance of the diameter distribution is presented in Fig. 2(b). The variance is found to increase linearly with the number of graphene layers, in agreement with the theoretical prediction. The best fit slope is 0.71 nm$^2$, about a factor of 2 below the prediction of the theory, and there is a constant offset of 3.67 nm$^2$ that we ascribe to uncertainty associated with the HRSEM measurement of the particle diameter.

In conclusion, we report a new method for synthesizing metal nanoparticles of selected size. This involves physical vapor deposition of metal onto graphene on $SiO_2$/Si wafers, followed by aging toward equilibrium in forming gas. Our modeling suggests that this method utilizes the differential electrostatic energy of graphene due to its strong thickness-dependent interlayer charge screening arising from the relativistic low energy carrier characteristic. The repulsion of dipoles induced in the nanoparticles limits the size of the formed nanoparticles. Measurements of the average nanoparticle diameter and the variation of this quantity are in very good agreement with theoretical predictions that the optimum nanoparticle size follows a simple 1/3 power law scaling rule with graphene thickness, and a linear increase in the variance of the size distribution. This method could be adapted for the growth of nanoparticles with controlled dimensions suitable for application in electronics, optical and magnetic devices, catalysts, and nanometrology.

This work was supported by JSTO/DTRA/ARO Grant # W911NF-06-1-0462. L.A.S. and T.L. are supported by NSF Grant DMR08-05136, and E.J.M. is supported by Department of Energy under Grant DE-FG02-ER45118. N.J.K. gratefully acknowledges support from the REU program of the Laboratory for Research on the Structure of Matter, NSF REU Site Grant DMR06-48953.

## References

1. Novoselov, K. S.; Geim, A. K.; Morozov, S. V.; Jiang, D.; Zhang, Y.; Dubonos, S. V.; Grigorieva, I. V.; Firsov, A. A., Electric Field Effect in Atomically Thin Carbon Films. *Science* **2004,** *306* (5696), 666-669.
2. Guinea, F., Charge distribution and screening in layered graphene systems. *Phys. Rev. B: Condens. Matter Mater. Phys.* **2007,** *75* (23), 235433/1-235433/7.
3. McCann, E., Asymmetry gap in the electronic band structure of bilayer graphene. *Physical Review B* **2006,** *74* (16), 161403(R).
4. Adam, S.; Hwang, E. H.; Galitski, V. M.; Das Sarma, S., A self-consistent theory for graphene transport. *Proc. Natl. Acad. Sci.* **2007,** *104* (47), 18392-18397.
5. Zhou, S. Y.; Gweon, G. H.; Fedorov, A. V.; First, P. N.; de Heer, W. A.; Lee, D. H.; Guinea, F.; Castro Neto, A. H.; Lanzara, A., Substrate-induced bandgap opening in epitaxial graphene. *Nat. Mater.* **2007,** *6* (10), 770-775.
6. Katsnelson, M. I.; Geim, A. K., Electron scattering on microscopic corrugations in graphene. *Philos. Trans. R. Soc.,* **2008,** *366* (1863), 195-204.
7. Ohta, T.; Bostwick, A.; Seyller, T.; Horn, K.; Rotenberg, E., Controlling the Electronic Structure of Bilayer Graphene. *Science* **2006,** *313* (5789), 951-954.
8. Datta, S. S.; Strachan, D. R.; Mele, E. J.; Johnson, A. T., Surface potentials and layer charge distributions in few-layer graphene films. *Nano Letters* **2009,** *9*, 7-11.
9. Penner, R. M., Mesoscopic Metal Particles and Wires by Electrodeposition. *J. Phys. Chem. B* **2002,** *106* (13), 3339-3353.
10. Daniel, M. C.; Astruc, D., Gold Nanoparticles: Assembly, Supramolecular Chemistry, Quantum-Size-Related Properties, and Applications toward Biology, Catalysis, and Nanotechnology. *Chem. Rev.* **2004,** *104* (1), 293-346.
11. Ferrari, A. C.; Meyer, J. C.; Scardaci, V.; Casiraghi, C.; Lazzeri, M.; Mauri, F.; Piscanec, S.; Jiang, D.; Novoselov, K. S.; Roth, S.; Geim, A. K., Raman spectrum of graphene and graphene layers. *Physical Review Letters* **2006,** *97*, 187401.
12. Graf, D.; Molitor, F.; Ensslin, K.; Stampfer, C.; Jungen, A.; Hierold, C.; Wirtz, L., Spatially Resolved Raman Spectroscopy of Single- and Few-Layer Graphene. *Nano Letters* **2007,** *7*, 238-242.
13. Cross, C. E.; Hemminger, J. C.; Penner, R. M., Physical Vapor Deposition of One-Dimensional Nanoparticle Arrays on Graphite: Seeding the Electrodeposition of Gold Nanowires. *Langmuir* **2007,** *23* (20), 10372-10379.
14. Evans, E. L.; Bahl, O. P.; Thomas, J. M., The decoration of and epitaxial growth of gold on graphite surfaces. *Carbon* **1967,** *5* (6), 587-9.
15. Ratke, L.; Voorhees, P. W., *Growth and Coarsening*. Springer: 2002.
16. Filleter, T.; Emtsev, K. V.; Seyller, T.; Bennewitz, R., Local work function measurements of epitaxial graphene. *Applied Physics Letters* **2008,** *93*, 133117.
17. Shan, B.; Cho, K., First Principles Study of Work Functions of Single Wall Carbon Nanotubes. *Physical Review Letters* **2005,** *94* (23), 236602.
18. Tyson, W. R.; Miller, W. A., Surface free energies of solid metals: Estimation from liquid surface tension measurements. *Surface Science* **1977,** *62*, 267-276.
19. Ashcroft, N. W.; Mermin, N. D., *Solid State Physics*. Brooks Cole: 1976.

**FIG. 1.** HRSEM images of Au nanoparticles grown on graphite with thickness of 1.0 μm and a single layer graphene flake. Inset in Fig. 1(a) indicates the histogram of nanoparticle edge orientations.

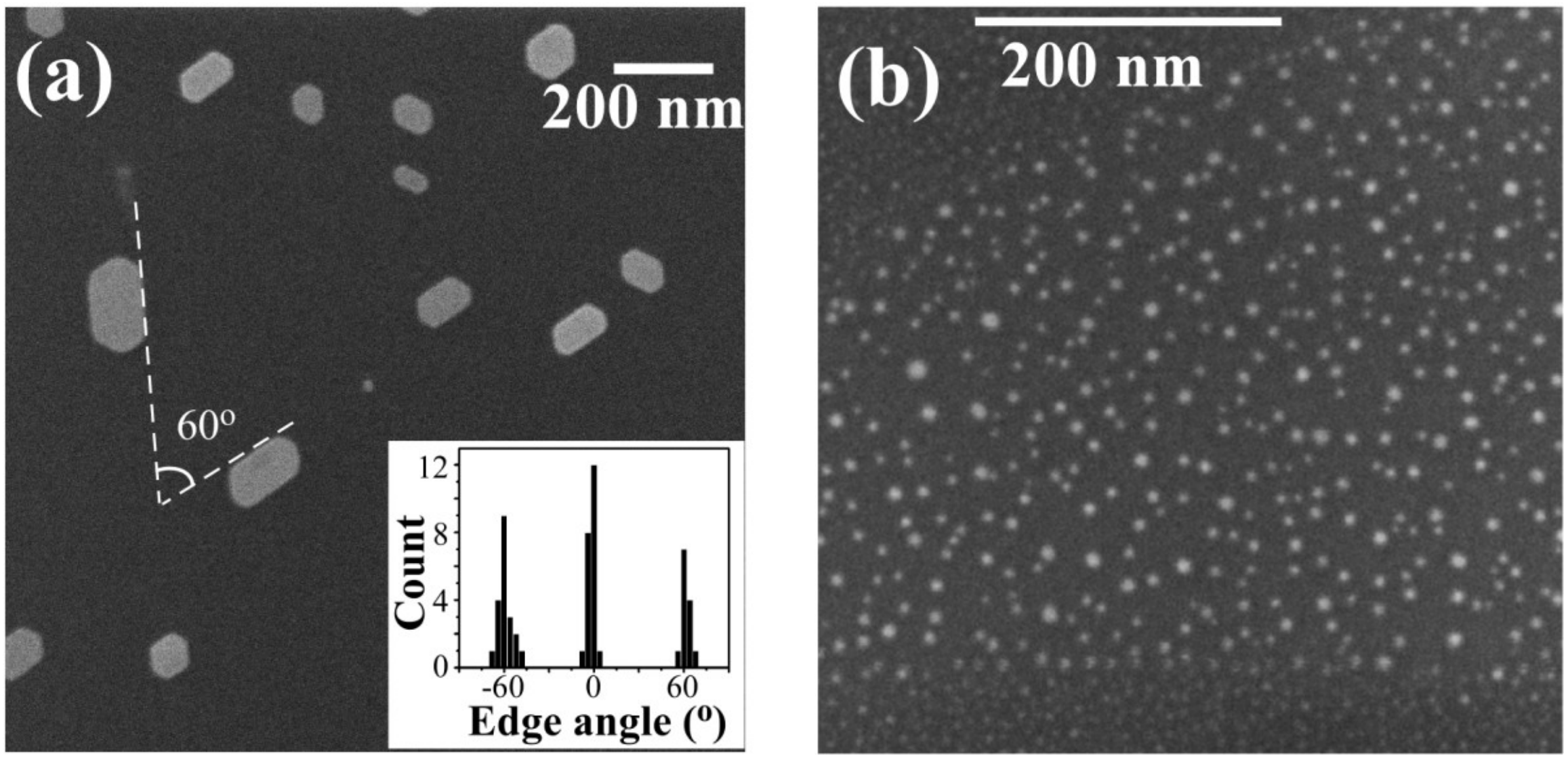

**FIG. 2.** (**a**) HRSEM images of Au nanoparticles prepared on few layer graphene on $SiO_2$/Si substrate. The numbers and arrow indicate the number of graphene layers. Inset in (**a**) is the AFM image of the same region before nanoparticle growth. (**b**) Average diameter of Au nanoparticles as a function of the number of graphene layers. The red line is a power law fit suggested by theory. Inset in **(b)** is the corresponding dependence of the variance of the diameter distribution; the curve shows the theoretically predicated linear dependence. The data point for 9 layers is shown in the graph but omitted from the fit.

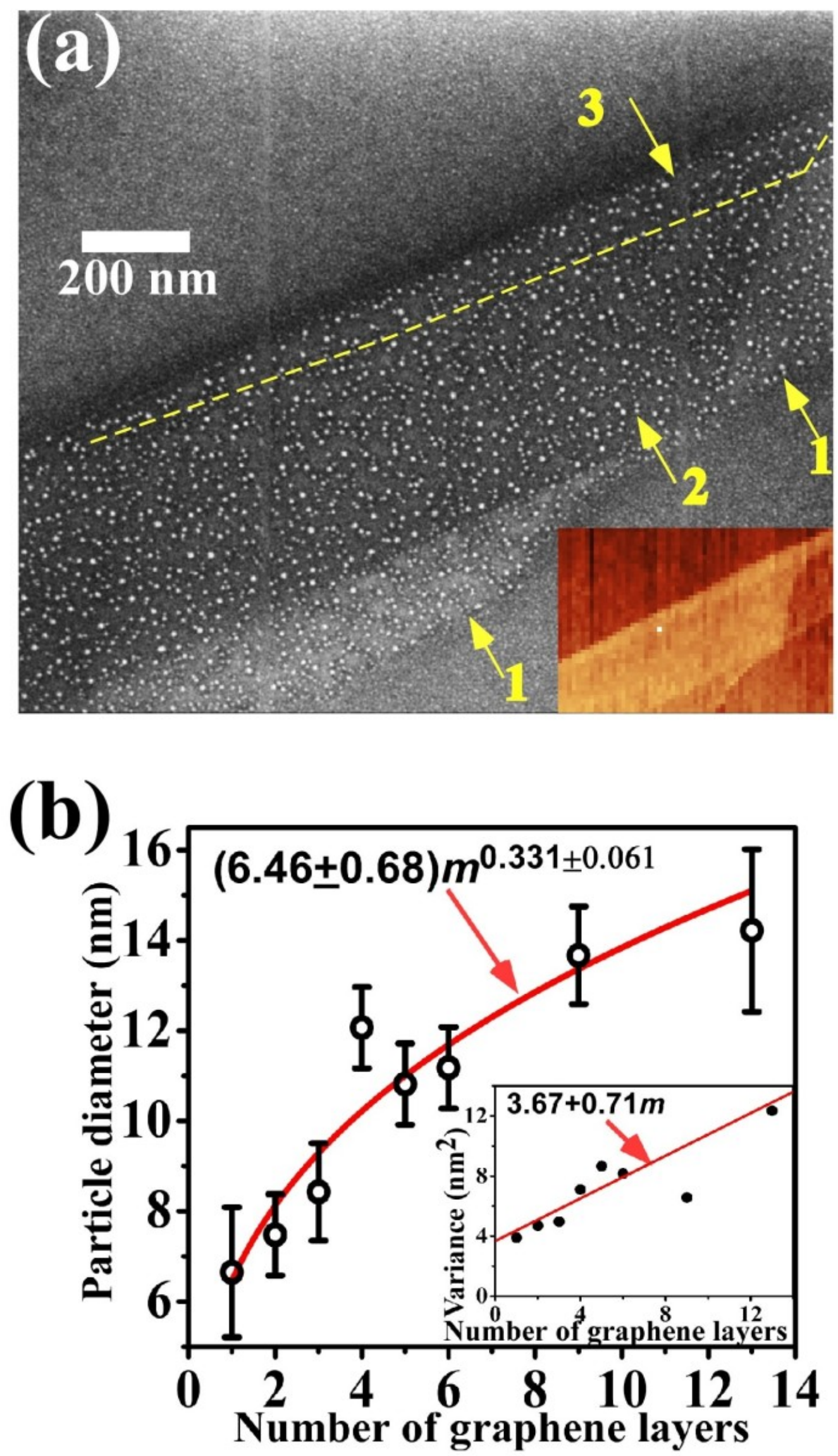

**FIG. 3.** Plot of the height of individual Au nanoparticles as a function of their lateral diameters. The inset shows a schematic of the sample.

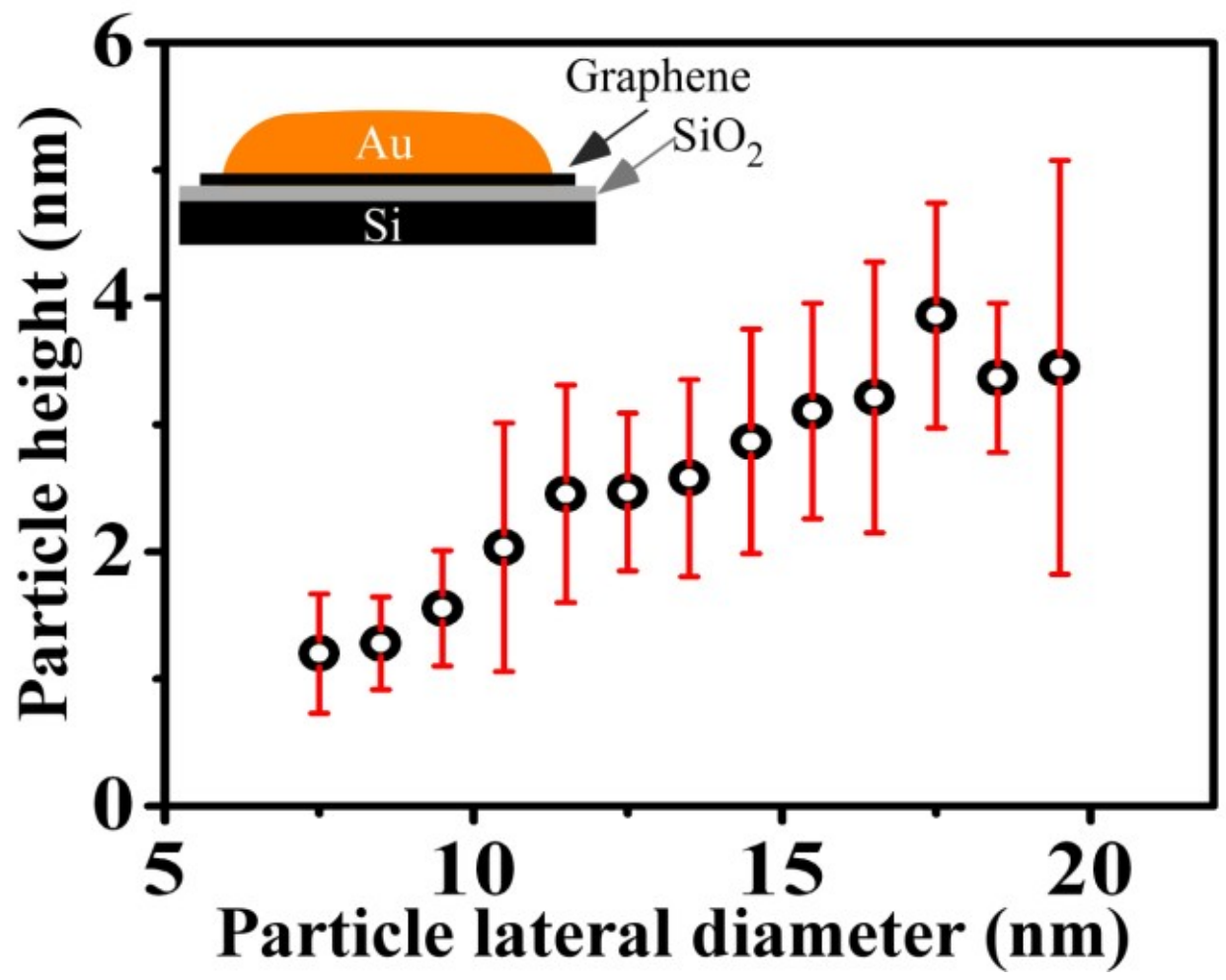

**FIG. 4.** (**a**) Plot of the phenomenological free energy as a function of disk diameter for various graphene thicknesses. The inset is an expanded view of the region in the dashed rectangle. (b) A plot of the chemical potential as a function of the disk diameter for graphene of various layer numbers. The inset is a schematic view of the system studied.

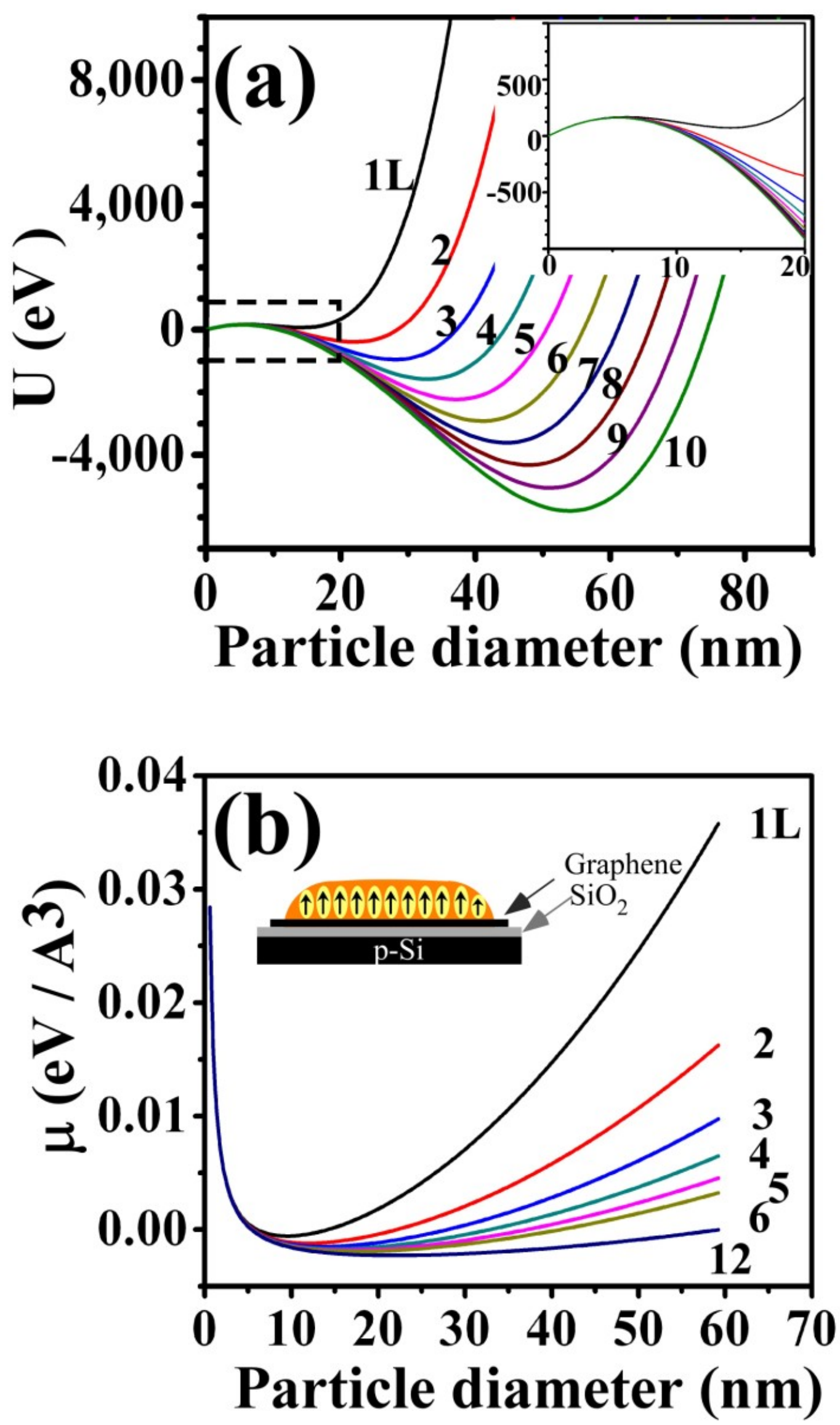